\title{HASN: Hybrid Attention Separable Network for Efficient Image Super-resolution}

\documentclass[pdflatex,sn-basic]{sn-jnl}

\usepackage{graphicx}%
\usepackage{picinpar}
\usepackage{wrapfig}
\usepackage{lipsum} 

\usepackage{subfigure}
\usepackage{multirow}%
\usepackage{amsmath,amssymb,amsfonts}%
\usepackage{amsthm}%
\usepackage[title]{appendix}%
\usepackage{xcolor}%
\usepackage{textcomp}%
\usepackage{manyfoot}%
\usepackage{booktabs}%
\usepackage{algorithm}%
\usepackage{algorithmicx}%
\usepackage{algpseudocode}%
\usepackage{listings}%
\usepackage{pifont}%
\usepackage{orcidlink}
\usepackage{lmodern}
\usepackage{scalefnt}

\usepackage{bbding}
\usepackage{color}

\begin{document}

\title{HASN: Hybrid Attention Separable Network for Efficient Image Super-resolution}

\author[1]{\fnm{Weifeng} \sur{Cao}}\email{weifeng\_cao@163.com}

\author*[1]{\fnm{Xiaoyan} \sur{Lei~\orcidlink{0009-0008-6328-799X}}}\email{xyan\_lei@163.com}

\author[1]{\fnm{Jun} \sur{Shi}}\email{shijunzz@gmail.com}

\author[1]{\fnm{Wanyong} \sur{Liang}}\email{lwy307@126.com}

\author[1]{\fnm{Jie} \sur{Liu}}\email{ljie35206609@163.com}

\author[1]{\fnm{Zongfei} \sur{Bai}}\email{zongfeibai@163.com}

\affil[1]{\orgdiv{the School of Electrical and Information Engineering}, \orgname{Zhengzhou University of Light Industry}, \orgaddress{\street{No.5 Dongfeng Road}, \city{Zhengzhou}, \postcode{450002}, \state{Henan}, \country{China}}}

\abstract{Recently, lightweight methods for single image super-resolution (SISR) have gained significant popularity and achieved impressive performance due to limited hardware resources. These methods demonstrate that adopting residual feature distillation is an effective way to enhance performance. However, we find that using residual connections after each block increases the model's storage and computational cost. Therefore, to simplify the network structure and learn higher-level features and relationships between features, we use depthwise separable convolutions, fully connected layers, and activation functions as the basic feature extraction modules. This significantly reduces computational load and the number of parameters while maintaining strong feature extraction capabilities. To further enhance model performance, we propose the Hybrid Attention Separable Block (HASB), which combines channel attention and spatial attention, thus making use of their complementary advantages. Additionally, we use depthwise separable convolutions instead of standard convolutions, significantly reducing the computational load and the number of parameters while maintaining strong feature extraction capabilities. During the training phase, we also adopt a warm-start retraining strategy to exploit the potential of the model further. Extensive experiments demonstrate the effectiveness of our approach. Our method achieves a smaller model size and reduced computational complexity without compromising performance. Code can be available at \url{https://github.com/nathan66666/HASN.git}}

\keywords{Efficient super-resolution, Channel Attention, Spacial Attention, Hybrid Attention Separable Block}

\maketitle

\section{Introduction}\label{sec1}

\begin{figure}[htbp]
\centering
\subfigure[PSNR results v.s the total number of parameters of different methods for image SR on Set5.]{
\begin{minipage}[t]{0.5\linewidth}
\centering
\includegraphics[width=2.5in]{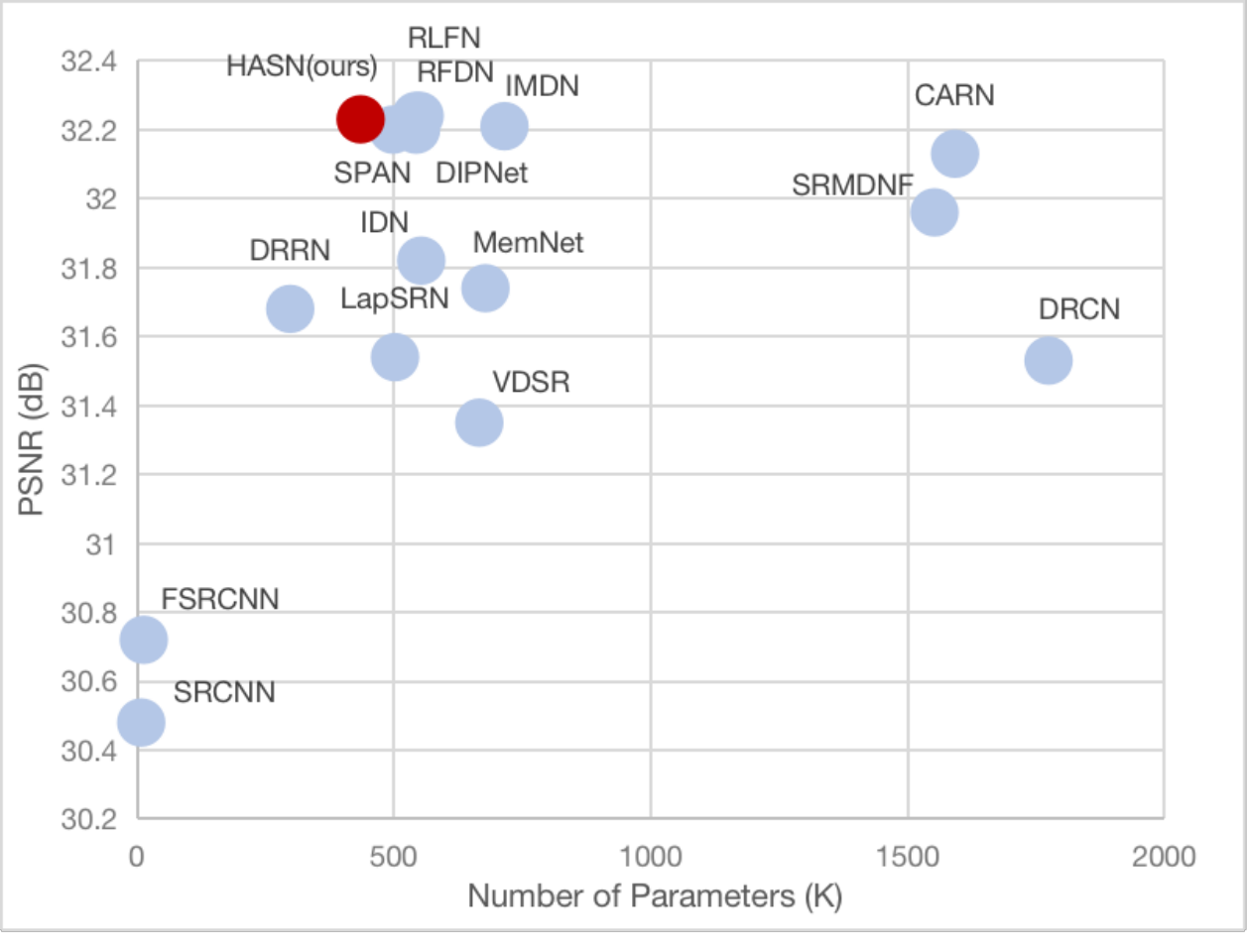}
\end{minipage}%
}%
\subfigure[SSIM results v.s the total number of parameters of different methods for image SR on Set5.]{
\begin{minipage}[t]{0.5\linewidth}
\centering
\includegraphics[width=2.5in]{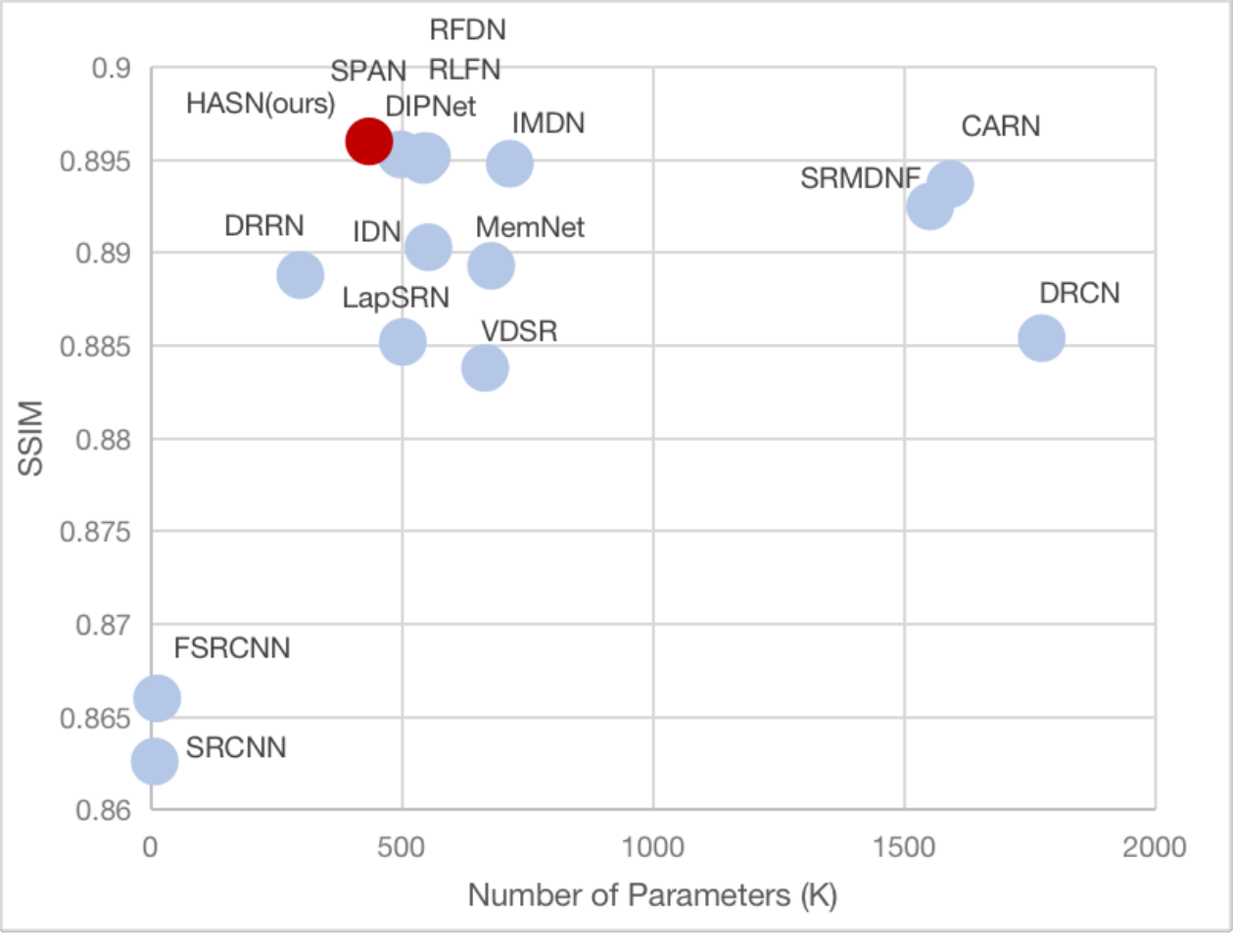}
\end{minipage}%
}%
\centering
\label{fig:table}
\caption{Comparison with other SOTA methods for image SR on Set5. The red dots represent the method proposed in this paper.}
\end{figure}
As the application scenarios of virtual reality technology continue to expand, so too does the demand for image quality. High-quality images can provide users with a more immersive experience. In this context, events such as the CGI and CASA conferences are dedicated to advancing various fields within computer graphics and virtual reality, making significant contributions to the progress of these technologies. The successful application of image super-resolution techniques will undoubtedly further promote the development of this field. Particularly, the emergence of efficient image super-resolution technology has made it easier to deploy this technology on edge devices, thereby broadening its application.

Image super-resolution (SR) is a typical branch of low-level vision methods, reconstructing high-resolution (HR) images from low-resolution (LR) inputs. Traditional SISR methods use interpolation techniques to recover corresponding HR images from LR ones. While simple and effective, these methods struggle to restore some of the details and textures in images. Since SRCNN~\cite{SRCNN} first introduced convolutional neural networks to the field of image super-resolution, deep learning (DL) has achieved remarkable performance and realistic visual effects due to its learnable feature representations. These SR networks~\cite{SwinIR, chen2021attention, dong2014learning, zhang2017learning, kim2016accurate, lim2017enhanced, zhang2018image, niu2020single, zhang2019residual, wang2018esrgan} have significantly improved the quality of reconstructed images. Their success can be partially attributed to their larger model capacity and intensive computational power. However, this makes them difficult to deploy on resource-constrained devices in real-world applications. Therefore, it is necessary to design lightweight models to improve the efficiency of SISR models, achieving a good balance between image quality and inference time.

Many prior works~\cite{SRCNN, FSRCNN, VDSR, DRCN, LapSRN, DRRN, MemNet, IDN, SRMDNF, CARN, IMDN, RFDN, RLFN, dipnet, SPAN} have been proposed to develop efficient image super-resolution models. They use different strategies to achieve high efficiency,
including parameter sharing strategy~\cite{kim2016deeply}, cascading network with grouped convolution~\cite{ahn2018fast}, information or feature distillation mechanisms~\cite{IMDN, RFDN, RLFN} and attention mechanisms~\cite{chen2021attention, RFDN, SwinIR}. Although they have improved efficiency using these strategies, redundancy still exists in convolution operations.

In this paper, to make the network more lightweight, we propose a new lightweight SR network, which consists of several stacked Hybrid Attention Separable Blocks. This structure is capable of extracting higher-level image features and includes more edge features and texture details. We only use a few necessary residual connections to prevent the vanishing gradient problem while integrating low-level features. Additionally, we use depthwise separable convolutions instead of standard convolutions in Convolutional Blocks, significantly reducing the computational load and the number of parameters while maintaining strong feature extraction capabilities. To fully maximize the model's capabilities, we propose a Warm-Start Retraining Strategy to further learn the image distribution, and use the Geometric Self-ensemble Strategy during the inference phase. Specifically, our contributions are as follows:

\begin{itemize}
\item We propose a hybrid attention separable network for efficient image super-resolution, which can extract higher-level image features and include more edge features and texture details without additional residual connections.

\item We propose a Warm-Start Retraining Strategy, which helps in learning the distribution of high-resolution images, effectively enhancing network performance.

\item Extensive experiments demonstrate that our proposed method surpasses existing state-of-the-art (SOTA) methods in terms of parameters and FLOPs, while maintaining comparable performance in PSNR and SSIM metrics.

\end{itemize}

\section{Related Work}
\subsection{Classical SISR methods}

SRCNN~\cite{SRCNN} is the first work that introduces deep convolutional neural networks (CNNs) to the image SR task. They use a three-layer convolutional neural network to jointly optimize feature extraction, nonlinear mapping, and image reconstruction in an end-to-end manner, achieving performance superior to traditional SR methods.
Subsequent methods adopt more complex convolutional module designs, such as residual blocks~\cite{RFANet, RFDN, blueprint} and dense blocks~\cite{RDN}, to enhance the model's representational capacity. 
As networks become larger and deeper, the introduction of various attention mechanisms~\cite{HAT, SwinIR} has become a new trend in image super-resolution research. For example, RCAN~\cite{RCAN} employs channel attention, while PAN~\cite{PAN} uses pixel attention. Additionally, self-attention mechanisms have shown significant performance in image reconstruction. SwinIR~\cite{SwinIR} leverages the Swin Transformer architecture~\cite{swint}, multi-scale feature representation~\cite{RepRFN}, hybrid attention mechanisms, and local-global feature interaction. HAT~\cite{HAT} further expands the window size and uses channel attention to better activate available pixels. PCCFormer~\cite{pccformer} use parallel attention transformer and adaptive convolution residual block to improve feature expression ability of the model. Recently, some emerging attention mechanisms have also achieved great success in imaging~\cite{lin2021eapt, zhou2022fsad}. Image super-resolution techniques have been applied in the medical field, making significant contributions to the diagnosis of brain diseases and morphometric studies~\cite{huang2023transmrsr}.

\subsection{Lightweight SISR methods}

To meet the requirements of edge devices, it is crucial to develop lightweight and efficient SR models. The SR network SRCNN~\cite{SRCNN} achieves impressive results but also faces issues such as high computational demands. FSRCNN~\cite{FSRCNN} addresses these issues by removing the interpolation upsampling, introducing transposed convolution at the end of the network, and using smaller but more numerous convolutional kernels, achieving approximately 17 times the acceleration compared to SRCNN. DRCN~\cite{DRCN} employs recursive calls to the feature extraction layers, while DRRN~\cite{DRRN} improves upon DRCN by combining recursive and residual networks to achieve better performance with fewer parameters. LapSRN~\cite{LapSRN} uses transposed convolution for upsampling, leveraging convolutional layers to learn the residuals between high-resolution images and upsampled feature maps, achieving multi-scale reconstruction through progressive upsampling. IDN~\cite{IDN} effectively extracts local long-path and short-path features through an information distillation module, achieving relatively fast inference speed. IMDN~\cite{IMDN} constructed a cable information multi-distillation block (IMDB) consisting of distillation and selective fusion. The distillation module gradually extracts features, while the fusion module determines the importance of candidate features based on an attention mechanism and fuses them accordingly.

Recently, researchers have been optimizing convolution methods to develop lighter and more efficient SR models. For example, ECBSR~\cite{ECBSR} and RepVGG~\cite{repvgg} effectively extract edge and texture information, while FMEN~\cite{FMEN} and BSRN~\cite{blueprint} further accelerate network inference and reduce the number of network parameters, achieving efficient super-resolution.

\section{Methodology}

\begin{figure}[!htbp]
\centering
\includegraphics[width=5in]{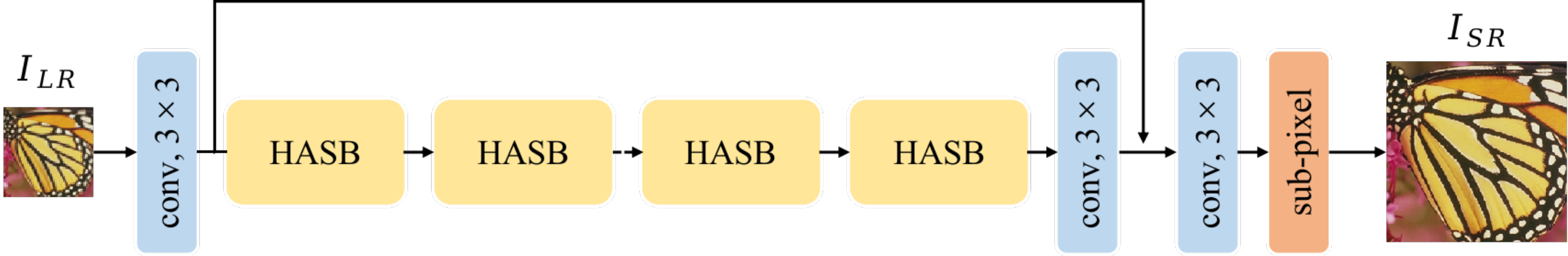}
\caption{The overall network architecture of our HASN.}
\label{fig:hasn}
\end{figure}

\subsection{Overall network architecture}

For the overall network structure of HASN, we adopt a coarse-to-fine strategy to learn representative features from LR images. As shown in~\ref{fig:hasn},  HASN consists of three main stages: an initial feature extraction, a multi-stage feature extraction, and a high-resolution reconstruction. Here, $I_{LR}$ represents the original image input, $I_{LR}\in \mathbb{R} ^{H\times W\times C_{in}}$((H, W and C are the image height, width and input channel number, respectively). A $3\times3$ convolutional layer $H_{IF}(\cdot)$ is used to extract
initial feature. This process can be expressed as:
\begin{equation} \label{eq1}
  \begin{split}
   F_{0} = H_{IF}(I_{LQ}), 
  \end{split}
\end{equation}
The convolutional layer effectively captures local features of an image, providing feature maps for subsequent deep feature extraction. Next, $F_0$ extracts multi-stage features using HASBs. we extract deep feature as:
\begin{equation} \label{eq2}
  \begin{split}
   & F_{i} = H_{{HASB}_{i}}(F_{0}) , i = 1, 2, \ldots, K,\\
   & F_{DF} = H_{Conv}(F_{K}), 
  \end{split}
\end{equation}
where $H_{{HASB}_{i}}(\cdot)$ denotes the $i-th$ HASB. A $3\times3$ convolutional layer is used after several HASBs to further process and refine the feature representations, enhancing the feature learning capability.

\begin{equation} \label{eq3}
  \begin{split}
   & I_{RHQ} = H_{REC}(F_{DF} + F_0), 
  \end{split}
\end{equation}
where $H_{REC}(\cdot)$ is the function of the reconstruction module. It consists of a \(3 \times 3\) convolutional layer and a sub-pixel layer. The \(3 \times 3\) convolutional layer reduces the dimensionality of the high-dimensional feature maps while preserving important information, preparing them for the sub-pixel layer. The entire training process is divided into two stages. The $\mathcal{L} _1$ loss function is exploited to optimize the model in the first stage, which can be formulated as follows:
\begin{equation} \label{eq4}
  \begin{split}
   & \mathcal{L} _1 = \left \| I_{SR}-I_{HR} \right \| _1, 
  \end{split}
\end{equation}
The loss function for the second stage$(\mathcal{L} _{s2})$ is defined as follows:

\begin{equation} \label{eq5}
  \begin{split}
   & \mathcal{L} _{s2} = \alpha \mathcal{L} _1 + \beta \mathcal{L} _{D_{KL}}, \\
   & \mathcal{L} _{D_{KL}} = \textstyle \sum_{i}^{}  P_{I_{HR}}(i)log\frac{P_{I_{HR}}(i)}{P_{I_{SR}}(i)} ,
  \end{split}
\end{equation}
where $\mathcal{L} _{D_{KL}}$ is KL divergence loss, which is used to measure the difference between the probability distributions of the actual high-resolution image and the predicted super-resolution image. $P_{I_{HR}}(i)$ represents the probability distribution of the $i-th$ pixel in the high-resolution image, and $P_{I_{SR}}(i)$ represents the probability distribution of the $i-th$ pixel in the super-resolution image.$ \alpha$ and $\beta$ are two different constants, which we set to 1 in this context.

\subsection{Hybrid Attention Separable Block}

\begin{figure}[!htbp]
\centering
\includegraphics[width=5in]{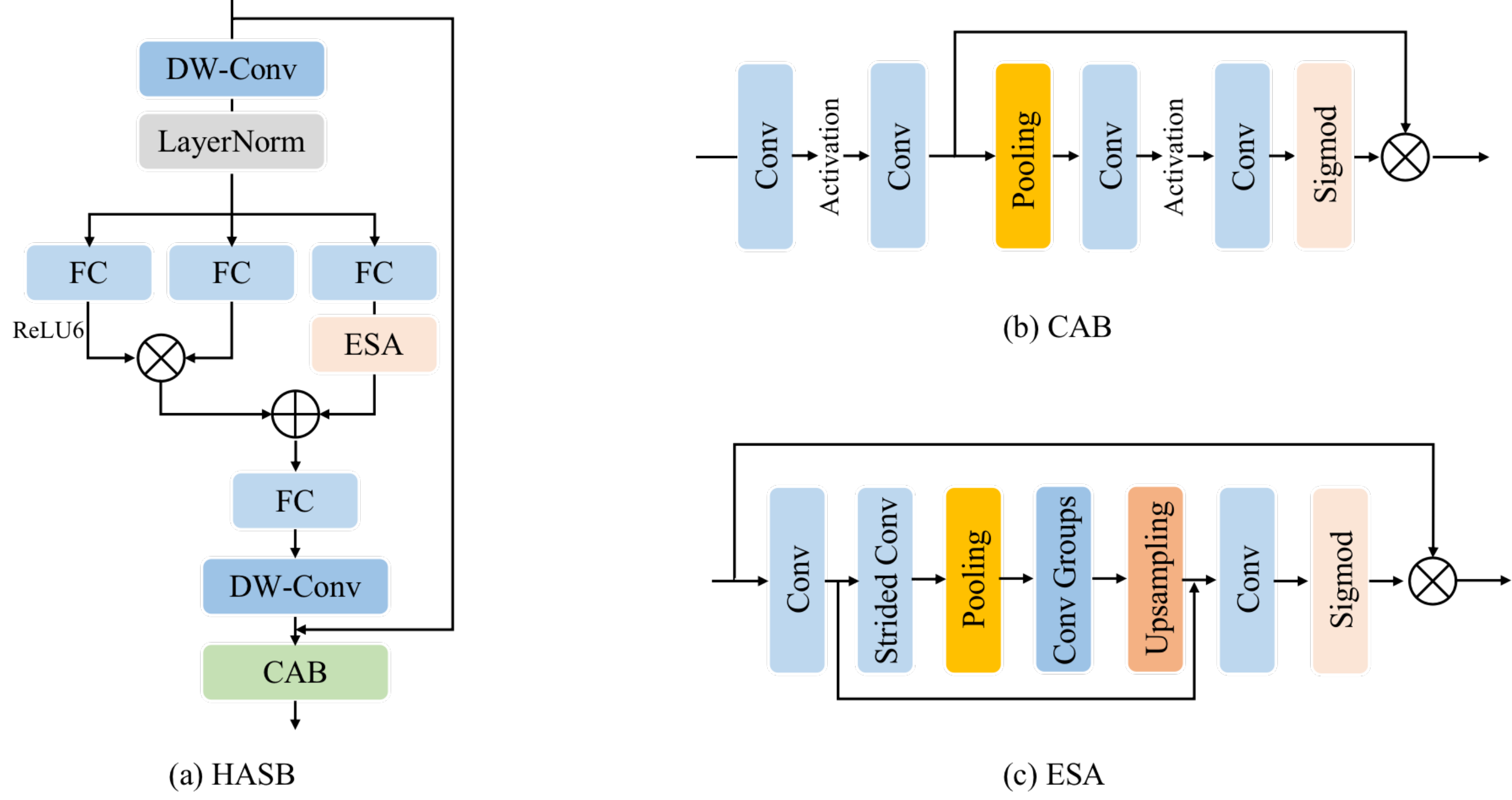}
\caption{(a) The architecture of Hybrid Attention Separable Block(HASB). (b)The architecture of Channel Attention Block(CAB). (c) The architecture of Enhanced Spatial Attention(ESA).}
\label{fig:hasb}
\end{figure}
As shown in Figure~\ref{fig:hasb}, our proposed HASB consists of two depthwise separable convolutions, several fully connected layers, a Channel Attention Block, and Enhanced Spatial Attention. First, a \(7 \times 7\) depthwise separable convolution operation is applied to the input features \(F_{in}\) to extract local features. Then, the convolved features are subjected to layer normalization, resulting in the normalized features \(F_{o}\). The normalized features \(F_{o}\) is passed to three parallel fully connected layers. The output of the first fully connected layer is passed through a ReLU6 activation function. The output of the second fully connected layer is used directly. The output of the third fully connected layer is processed through the Enhanced Spatial Attention module. The output of the first fully connected layer is multiplied element-wise with the output of the second fully connected layer. The result of this multiplication is added element-wise to the output of the third fully connected layer (features processed by the ESA) to obtain the fused features. The fused features are passed to a fully connected layer for further processing. The features processed by the fully connected layer are passed through another depthwise separable convolution layer to extract additional features. Finally, the features are processed through the Channel Attention Block module to obtain the final output features. The input feature \(F_{in}\) is added directly to the features before the final depthwise separable convolution layer (DW-Conv) through a residual connection. This design helps alleviate the vanishing gradient problem and enhances feature learning. The whole structure is described as
\begin{equation} \label{eq6}
  \begin{split}
   & F_{o} = LN(DWConv_{7\times 7}(F_{in} )), \\
   & F_{d_1}, F_{d_2}, F_{d_3} = FC(F_{o}), FC(F_{o}), FC(F_{o}),\\
   & F_{d} = ReLU6(F_{d_1})\otimes F_{d_2} + ESA(F_{d_3}), \\
   & F_{d} = DWConv_{7\times 7}(FC(F_{d})) + F_{in}), \\
   & F_{out} = CAB(F_{d})
  \end{split}
\end{equation}
where \(DWConv_{7\times 7}\) represents a depthwise separable convolution with a \(7 \times 7\) kernel, \(LN(\cdot)\) denotes the LayerNorm layer, and FC refers to the fully connected layer.

\subsection{Warm-Start Retraining Strategy}

We propose a novel warm-start retraining strategy. Different from some previous works that use the \(2\times\) model as a pre-trained network instead of training from scratch, we train HASN for \(4\times\) from scratch in the first stage. In the second stage, we load the model weights from the first stage, which are not fully converged, and further expand the dataset (adding Flickr2K). We further learn the distribution of high-resolution images by minimizing the KL divergence loss and L1 loss, as formulated in Equation~\ref{eq5}. The other training settings remain consistent with the first stage.

\section{Experiments}

\subsection{Datasets and metrics}
In this paper, the entire training process is divided into two stages. In the first stage, we use the DIV2K~\cite{div2k} dataset, and in the second stage, we use the DF2K dataset (DIV2K + Flickr2K)~\cite{div2k} to further improve the network performance. DIV2K~\cite{div2k} is a high-quality (2K resolution) image dataset containing 800 training images. Flickr2K is an image dataset with 2K resolution containing 2,650 images. Additionally, the low-resolution images of DIV2K and Flickr2K are generated from the ground truth images by the ``bicubic'' downsampling in MATLAB. For testing, we use five widely-used benchmark datasets: Set5~\cite{Set5}, Set14~\cite{Set14}, BSD100~\cite{B100}, Urban100~\cite{Urban100}, and Manga109~\cite{Manga109}. We evaluate all the SR results using the PSNR and SSIM metrics on the Y channel of the YCbCr color space.

\begin{table*}[t]
    \centering
    \caption{Average PSNR/SSIM for scale factor 4 on datasets Set5, Set14, BSD100, Urban100, and Manga109. The best and second best results are highlighted in {\color{red}red} and {\color{blue}blue} respectively.}
    \resizebox{0.9\linewidth}{!}{
    \begin{tabular}{lcccccccc}
    \hline
    \multirow{2}{*}{Method}  & \multirow{2}{*}{Params} & \multirow{2}{*}{FLOPs(G)
} & Set5 & Set14 & BSD100 & Urban100 & Manga109 \\
    & \multicolumn{1}{l}{} & \multicolumn{1}{l}{}   & PSNR/SSIM & PSNR/SSIM & PSNR/SSIM & PSNR/SSIM & PSNR/SSIM \\
    \hline
    \hline
    Bicubic  & - & - & 28.42/0.8104 & 26.00/0.7027 & 25.96/0.6675 & 23.14/0.6577 & 24.89/0.7866 \\
    SRCNN~\cite{SRCNN}  & 8K & 52.7 & 30.48/0.8626 & 27.50/0.7513 & 26.90/0.7101 & 24.52/0.7221 & 27.58/0.8555 \\
    FSRCNN~\cite{FSRCNN}  & 13K & 4.6 & 30.72/0.8660 & 27.61/0.7550 & 26.98/0.7150 & 24.62/0.7280 & 27.90/0.8610 \\
    VDSR~\cite{VDSR}  & 666K & 612.6 & 31.35/0.8838 & 28.01/0.7674 & 27.29/0.7251 & 25.18/0.7524 & 28.83/0.8870 \\
    DRCN~\cite{DRCN}  & 1774K & 17,974.0 & 31.53/0.8854 & 28.02/0.7670 & 27.23/0.7233 & 25.14/0.7510 & 28.93/0.8854 \\
    LapSRN~\cite{LapSRN}  & 502K & 149.4 & 31.54/0.8852 & 28.09/0.7700 & 27.32/0.7275 & 25.21/0.7562 & 29.09/0.8900 \\
    DRRN~\cite{DRRN}  & 298K & 6,796.9 & 31.68/0.8888 & 28.21/0.7720 & 27.38/0.7284 & 25.44/0.7638 & 29.45/0.8946 \\
    MemNet~\cite{MemNet}  & 678K & 2662.4 & 31.74/0.8893 & 28.26/0.7723 & 27.40/0.7281 & 25.50/0.7630 & 29.42/0.8942 \\
    IDN~\cite{IDN}  & 553K & 81.8 & 31.82/0.8903 & 28.25/0.7730 & 27.41/0.7297 & 25.41/0.7632 & 29.41/0.8942 \\
    SRMDNF~\cite{SRMDNF}  & 1552K & 89.3 & 31.96/0.8925 & 28.35/0.7787 & 27.49/0.7337 & 25.68/0.7731 & 30.09/0.9024 \\
    CARN~\cite{CARN}  & 1592K & 90.9 & 32.13/0.8937 & 28.60/0.7806 & {27.58}/0.7349 & 26.07/0.7837 & 30.47/{0.9084} \\
    IMDN~\cite{IMDN}  & 715K & 40.9 & {32.21}/{0.8948} & 28.58/0.7811 & 27.56/0.7353 & 26.04/0.7838 & 30.45/0.9075 \\
    RFDN~\cite{RFDN} & 550K & {\color{blue}31.6} & 32.24/0.8952 & 28.61/{0.7819} & 27.57/0.7360 & 26.11/0.7858 & {\color{blue}30.58}/{\color{blue}0.9089} \\
    RLFN~\cite{RLFN}  & 543K & 33.9 & {\color{red}32.24}/{0.8952} & {\color{blue}28.62}/0.7813 & {\color{blue}27.60}/0.7364 & {\color{blue}26.17}/{\color{blue}0.7877} & -/- \\
    
    DIPNet~\cite{dipnet}  & 543K & - & 32.20/0.8950& {28.58}/0.7811 &{27.59}/{0.7364} &{26.16}/{0.7879}& {30.53}/{0.9087}\\
    
    SPAN~\cite{SPAN}  & 498K & - & {32.20}/{\color{blue}0.8953} & 28.66/{\color{red}0.7834} & 27.62/{\color{blue}0.7374 }& {\color{red}26.18}/{\color{red}0.7879} & {\color{red}30.66}/{\color{red}0.9103} \\
    
    HASN (Ours)  & 435K & {\color{red}26.6} & {\color{blue}32.23}/{\color{red}0.8960} &  {\color{red}28.66}/{\color{blue}0.7830}&  {\color{red}27.62}/{\color{red}0.7387} &  26.13/0.7869& 30.50/0.9077 \\
    \hline
    \end{tabular}}
    \label{tab:sota}
\end{table*}

\subsection{Implementation details}
The proposed HASN consists of 6  HASBs and the number of channels is set to 52. The kernel size of all depth-wise convolutions is set to 7. During training, we set the input patch size to 192 $\times$ 192 and use random rotation and horizontal flipping for data augmentation. The batch size is set to 128 and the total number of iterations is 500k. The initial learning rate is set to $2\times 10^{-4}$. We adopt a multi-step learning rate strategy, where the learning rate will be halved when the iteration reaches 250000, 400000, 450000, and 475000, respectively. The model is trained by Adam optimizer with $\beta _{1} = 0.9$ and $\beta _{2} = 0.99$. In the second stage of training, we chose the model weights from the 100k-th iteration of the first stage as the starting point, and the total number of iterations is set to 1000k. Additionally, we use \(\mathcal{L}_{s2}\) as the loss function for the second stage. Other training settings remain consistent with the first stage. To maximize the potential performance of the HASN proposed in this paper, we use Geometric Self-ensemble~\cite{lim2017enhanced} in the experiment, which is applied during inference without additional training. The networks are implemented by using PyTorch framework with a NVIDIA 3090 GPU.

\subsection{Comparison with state-of-the-arts}

We compare our models with several advanced efficient super-resolution models with scale factor of 4. The comparison methods include SRCNN~\cite{SRCNN}, FSRCNN~\cite{FSRCNN}, VDSR~\cite{VDSR}, DRCN~\cite{DRCN}, LapSRN~\cite{LapSRN}, DRRN~\cite{DRRN}, MemNet~\cite{MemNet}, IDN~\cite{IDN}, SRMDNF~\cite{SRMDNF}, CARN~\cite{CARN}, IMDN~\cite{IMDN}, RFDN~\cite{RFDN}, RLFN~\cite{RLFN}, DIPNet~\cite{dipnet}, SPAN~\cite{SPAN}. Firstly, in terms of model performance, we use PSNR and SSIM as evaluation metrics. In terms of model efficiency, we use Parameters and FLOPs to measure the model size and computational complexity. The quantitative performance comparison on five benchmark datasets is shown in Table~\ref{tab:sota}. Compared with other state-of-the-art models, it can be seen that HASN achieves better performance on Set5, Set14, and BSD100. Its performance on the remaining two datasets is comparable. Overall, HASN achieves performance comparable to other networks with fewer parameters and computational complexity, achieving a better balance in performance and efficiency.

\section{Ablation Study}
In this section, we conduct a set of ablation experiments to evaluate the performance of each proposed module.

\subsection{{The choice of multiplication and addition in convolution block}}
\begin{figure}[!htbp]
\centering
\includegraphics[width=6in]{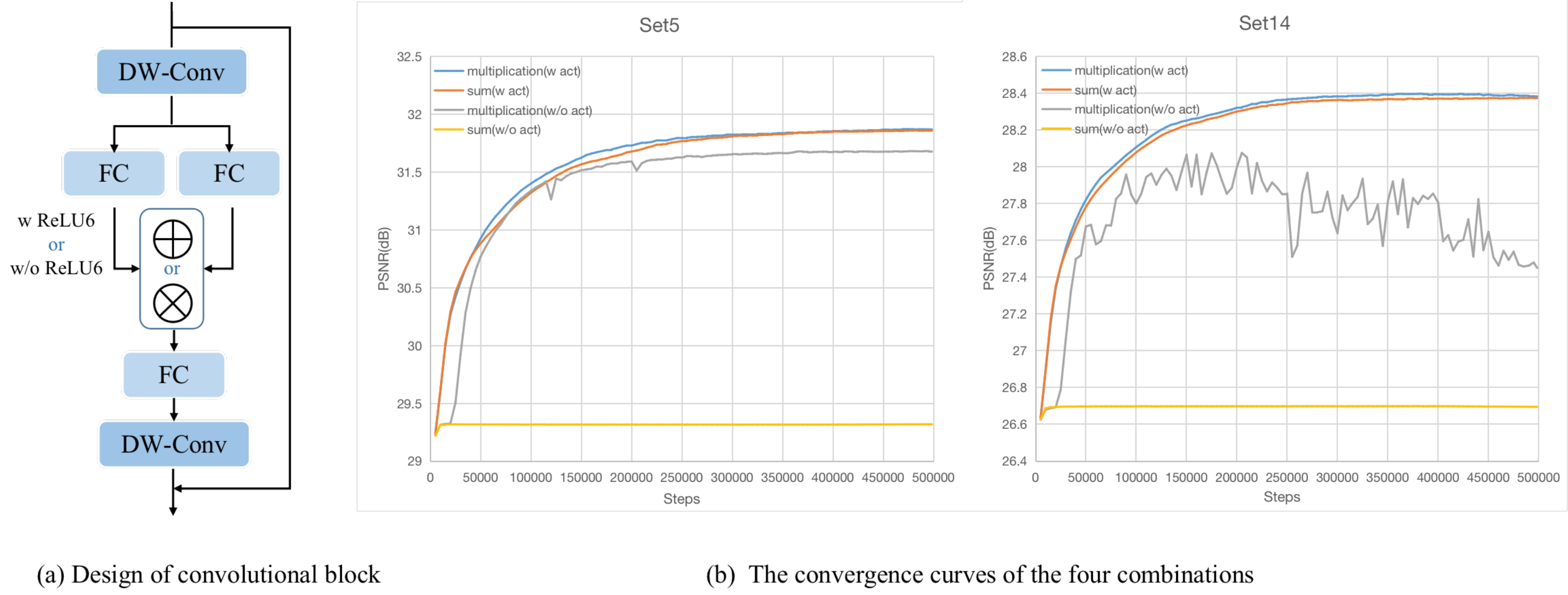}
\caption{{Design of convolutional block and convergence curves of different combinations.}}
\label{fig:motivation}
\end{figure}

Many previous efficient image SR methods~\cite{RFDN,SPAN,osffnet} benefit from residual connections, which extract features from each block up to the upsampling layer. Some methods~\cite{RFDN,RLFN,IMDN} also perform feature distillation within each block. However, these approaches often make the network structure redundant. We want to design an efficient and compact network. Inspired by~\cite{star}, element-wise multiplication seems to provide greater gains in a narrower network compared to addition. This finding is beneficial for our task, as we need to minimize network size while achieving equal or better performance compared to previous methods. Therefore, we design some simple experiments to validate this conclusion.
As shown in the figure~\ref{fig:motivation}, (a) presents the structure of the CB module. (b) illustrates the fitting curves of four different configurations. It is evident that when activation function is not used, element-wise multiplication performs significantly better than addition, despite some instability during training. When activation function is included, both addition and multiplication configurations exhibit smooth fitting curves, and the PSNR on the test set shows that the network using multiplication slightly outperforms the one using addition.
As shown in Table~\ref{tab:mul}, we set up networks with three different embedding dimensions. We find that in Urban100, the PSNR gain between element-wise multiplication and addition decreases as the dimension increases, from 0.08 dB to 0.07 dB, and finally to 0.01 dB. On other test sets, the changes do not seem to follow a consistent pattern. However, across various dimensions, using element-wise multiplication generally yields better performance.

\begin{table*}[!htbp]
\centering
\caption{Quantitative comparison (average PSNR/SSIM) of element-wise multiplication and addition across different embedding dimensions on benchmark datasets.}
\label{tab:mul}
\resizebox{\linewidth}{!}{
\begin{tabular}{ccccc|cc|cc|cc|cc|cc}
\toprule
\multirow{2}{*}{sum} & \multirow{2}{*}{multiplication} & \multirow{2}{*}{dim}& \multirow{2}{*}{param} & \multirow{2}{*}{FLOPs(G)} &\multicolumn{2}{c|}{Set5} & \multicolumn{2}{c|}{Set14} & \multicolumn{2}{c|}{B100} & \multicolumn{2}{c|}{Urban100} & \multicolumn{2}{c}{Manga109} \\ 
       &        &       &     &    & PSNR        & SSIM        & PSNR         & SSIM        & PSNR        & SSIM        & PSNR          & SSIM          & PSNR          & SSIM         \\ \hline
    \ding{51}   &    \ding{55}    &   30    &  90k  &   5.77  &   31.45    &   0.8847    &      28.13   &    0.7704   &    27.27    &   0.7272    &    25.15      &    0.7539     &    29.03     &    0.8868    \\

    \ding{55}   &   \ding{51}     &   30    &   90k   &  5.77    &   \textbf{31.52}  &   \textbf{0.8858}    &    \textbf{28.18}   &    \textbf{0.7716}     &    \textbf{27.30}   &    \textbf{0.7280}    &   \textbf{25.23}   &    \textbf{0.7560}      &   \textbf{29.04}     &     \textbf{0.8871}            \\\hline

    \ding{51}   &    \ding{55}    &   52    &  227k &  14.73    &    31.85   &   0.8908    &    28.36     &    0.7764   &    27.42    &   0.7328    &    25.52      &    0.7678     &    \textbf{29.67}     &   0.8979     \\

   \ding{55}   &   \ding{51}     &    52   &    227k   & 14.73     &    \textbf{31.87}   &  \textbf{0.8915}     &    \textbf{28.38}     &   \textbf{0.7770}    &    \textbf{27.45}    &   \textbf{0.7338}    &   \textbf{25.59}       &    \textbf{0.7700}     &     29.53    &    \textbf{0.8981}    \\\hline

    \ding{51}   &    \ding{55}    &   90    &  610k  &  39.62    &   32.01    &   0.8933    &    28.47     &   0.7799    &    27.52    &    0.7362   &     25.86     &   0.7795      &    \textbf{30.08}     &    \textbf{0.9039}    \\

        \ding{55}   &   \ding{51}     &   90    &    610k   &  39.62    &   \textbf{32.07}    &  \textbf{0.8940}     &     \textbf{28.51}    &   \textbf{0.7809}    &    \textbf{27.54}    &  \textbf{0.7368}     &   \textbf{25.87}       &    \textbf{0.7800}     &     29.91    &    0.9031    \\\bottomrule
\end{tabular}}
\end{table*}

\subsection{Study on HASB number}

\begin{figure}[!htbp]
\centering
\includegraphics[width=3in]{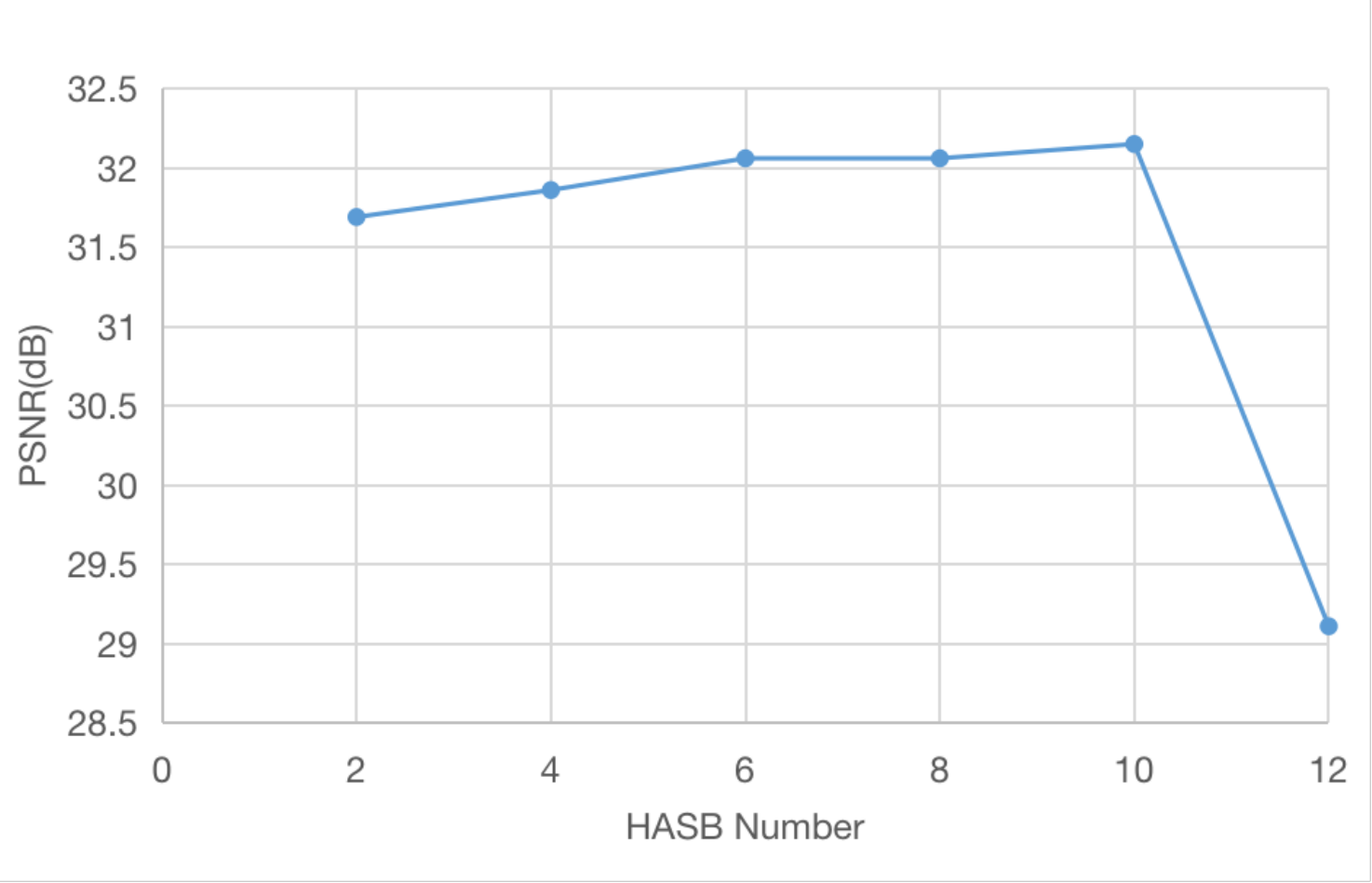}
\caption{{pink}{PSNR of different numbers of HASB on Set5.}}
\label{fig:hasb-num}
\end{figure}

From Figure~\ref{fig:hasb-num}, we can observe that with the increase in the number of HASBs, the PSNR shows an upward trend when the HASB number is less than or equal to 10. However, when the HASB number is set to 12, there is a sharp decline in PSNR for Set5. This phenomenon indicates that while increasing the number of HASB modules can enhance the model's feature extraction capability to some extent, excessively increasing them may lead to overfitting the training data. Due to the complexity of the attention mechanism and fully connected layers within the HASB modules, the model may capture noise and details from the training data, resulting in a reduced generalization ability on the test data. As shown in Table~\ref{tab:hasb-number}, with the increase in the number of HASBs, the model's parameter count and computational complexity also increase. Setting the HASB number to 6 balances the model size and performance.

\begin{table*}[!htbp]
\centering
\caption{{Quantitative comparison (average PSNR/SSIM) of different HASB number on benchmark datasets}}
\label{tab:hasb-number}
\resizebox{0.5\linewidth}{!}{
\begin{tabular}{c|c|c|cc|cc}
\toprule
\multirow{1}{*}{HASB}  & \multirow{2}{*}{param}& \multirow{2}{*}{FLOPs}& \multicolumn{2}{c|}{Set5} & \multicolumn{2}{c}{Set14}  \\ 
          Number           &          &   & PSNR        & SSIM        & PSNR         & SSIM        \\ \hline
    2                & 177k   & 10.98 &   31.69      & 0.8878      &    28.26     &     0.7742       \\

    4                & 306k     & 18.81  &  31.86    &   0.8909    &    28.40     &   0.7776      \\

    6                & 435k     & 26.63  &   32.06   &   0.8938    &    28.52     &    0.7803      \\
    
    8           &564k      & 34.46  &   32.06   &   0.8933    &    28.50     &     0.7785    \\ 
    10           &693k      & 42.28  &  32.15   &  0.8948    &    28.56     &    0.7808     \\
    12           &822k      & 50.11  &   29.11   &   0.8268    &    26.58     &    0.7277     \\ \bottomrule
\end{tabular}}
\end{table*}

\subsection{Study on kernel size of depth-wise convolution}

\begin{table*}[!htbp]
\centering
\caption{Quantitative comparison of different kernel sizes. We use the average PSNR/SSIM on the datasets Set5, Set14, BSD100, Urban100, and Manga109 as the metric. The best results are in bold.}
\label{tab:kernel_size}
\resizebox{\linewidth}{!}{
\begin{tabular}{c|c|c|cc|cc|cc|cc|cc}
\toprule
\multirow{2}{*}{kernel size} & \multirow{2}{*}{param}& \multirow{2}{*}{FLOPs}& \multicolumn{2}{c|}{Set5} & \multicolumn{2}{c|}{Set14} & \multicolumn{2}{c|}{B100} & \multicolumn{2}{c|}{Urban100} & \multicolumn{2}{c}{Manga109} \\ 
                     &          &   & PSNR        & SSIM        & PSNR         & SSIM        & PSNR        & SSIM        & PSNR          & SSIM          & PSNR          & SSIM         \\ \hline
    $3\times3$                 & 202k   &13.09   &31.74      & 0.8900      & 28.32        & 0.7759      & 27.39       & 0.7315      & 25.50        & 0.7656        & 29.49        &  0.8961      \\

    $5\times5$                 & 212k     &13.75   &31.79      & 0.8903      & 28.36        & 0.7768      & 27.43       & 0.7329      & 25.56         & 0.7687        & 29.65        &  0.8979      \\

    $7\times7$                 & 227k     &14.73   &{31.87}      & \textbf{0.8915}      & 28.38        & 0.7770      & \textbf{{27.45}}       & \textbf{{0.7338}}      & \textbf{{25.59}}         & {0.7700}        & {29.53}        &  \textbf{{0.8981} }     \\

    $9\times9$            &247k      &16.04   &\textbf{31.89}      & {0.8915}      &\textbf{28.41}         &\textbf{0.7774}      & {27.45}      &0.7336       &25.58          &\textbf{0.7705}        &\textbf{29.61}         &0.8979       \\ \bottomrule
\end{tabular}}
\end{table*}

To explore the impact of convolution kernel size on network performance, we set the kernel sizes of all depth-wise convolutions to 3, 5, 7, and 9, respectively. As shown in Table~\ref{tab:kernel_size}, we observed that performance improves with larger kernel sizes across the five benchmark datasets. However, as the kernel size increases, the number of network parameters and FLOPs also increase. From the table, the best results are seen between kernel sizes 7 and 9. To balance computational complexity and the number of parameters, choosing a kernel size of 7 is appropriate.

\subsection{Study on residual connection}

\begin{figure}[!htbp]
\centering
\includegraphics[width=5in]{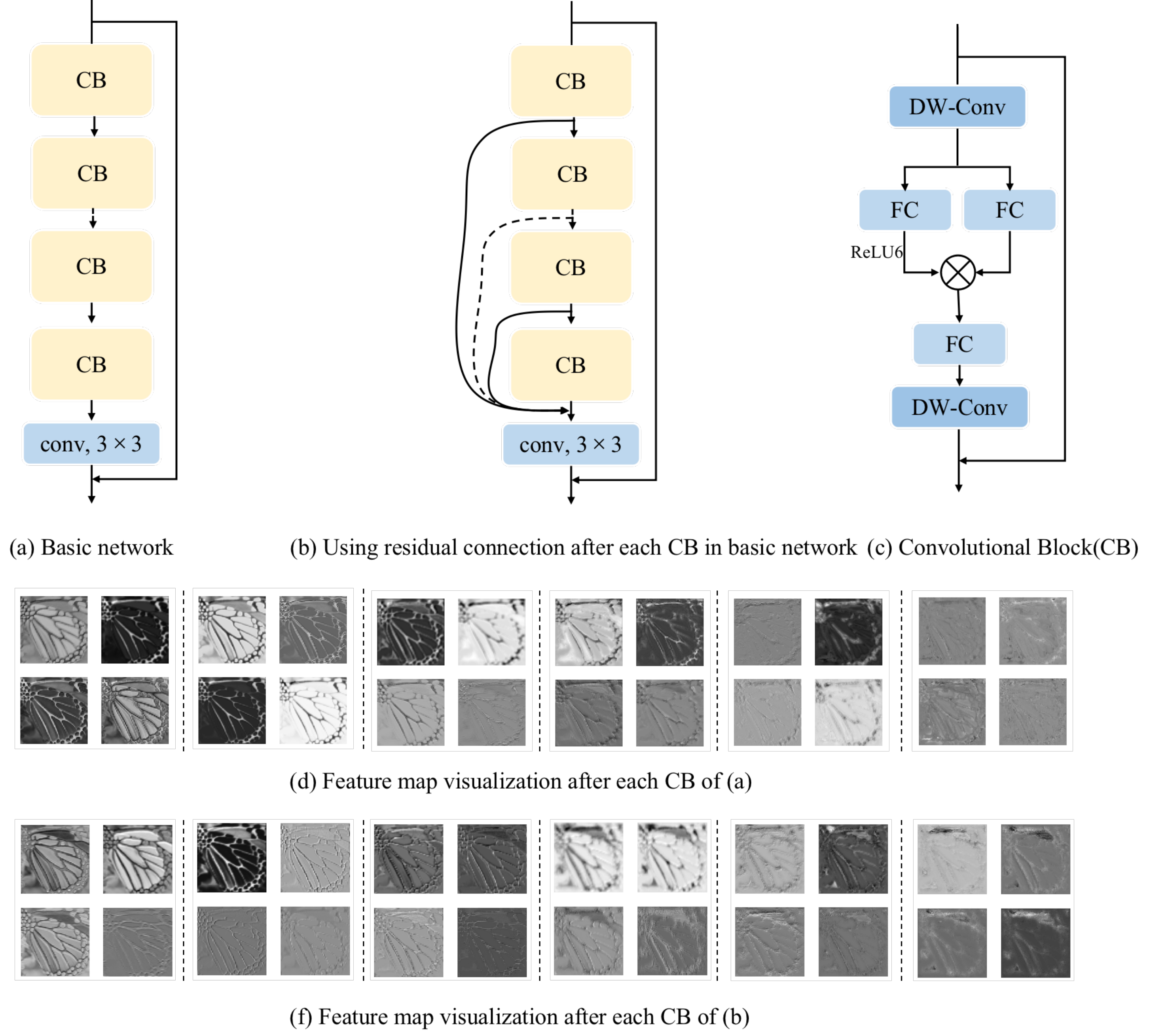}
\caption{(a) The basic network consists of several CBs and a \(3 \times 3\) convolutional layer. (b) Based on (a), a residual connection is used after each CB. (c) The network structure of the convolutional block. (d) Feature map visualization of the intermediate layers in (a) and (b).}
\label{fig:residual_connection}
\end{figure}

To explore the role of residual connections in image super-resolution, we use intermediate feature visualization to observe the changes in the network's intermediate features, as shown in Figure~\ref{fig:residual_connection}. (d) and (f) show feature map visualizations without and with residual connections, respectively. From left to right, the features progress from lower to higher layers, gradually shifting from capturing detailed information (such as edges and textures) to more abstract information (such as shapes and overall contours). The lower layer feature maps focus more on local features, while the information in the feature maps becomes more abstract and global as the layers deepen.

Comparing (d) and (f), we observe that the feature maps in (d) capture more information at each layer, retaining more edge and texture details. In contrast, the feature maps in (f) lose detail information more quickly and shift to more abstract representations. This suggests that in our method, CBs~\cite{star} may be sufficient to learn important features, while using excessive residual connections could introduce noise. The quantitative performance comparison on several benchmark datasets is shown in Table~\ref{tab:residual_connection}. The PSNR on Set5, Set14, B100, Urban100, and Manga109 improved by 0.13dB, 0.07dB, 0.03dB, 0.08dB, and 0.02dB, respectively.

\begin{table*}[!htbp]
\centering
\caption{Quantitative comparison of networks with and without residual connections. (a) represents the network without residual connections, and (b) represents the network with residual connections. We use the average PSNR/SSIM on the datasets Set5, Set14, BSD100, Urban100, and Manga109 as the metric. The best results are in bold.}
\label{tab:residual_connection}
\resizebox{\linewidth}{!}{
\begin{tabular}{c|cc|cc|cc|cc|cc}
\toprule
\multirow{2}{*}{Method} & \multicolumn{2}{c|}{Set5} & \multicolumn{2}{c|}{Set14} & \multicolumn{2}{c|}{B100} & \multicolumn{2}{c|}{Urban100} & \multicolumn{2}{c}{Manga109} \\ 
                               & PSNR        & SSIM        & PSNR         & SSIM        & PSNR        & SSIM        & PSNR          & SSIM          & PSNR          & SSIM         \\ \hline
    (a)                 &  \textbf{31.87}      & \textbf{0.8915}      & \textbf{28.38}        & \textbf{0.7770}      & \textbf{27.45}       & \textbf{0.7338}      & \textbf{25.59}         & \textbf{0.7700}        & \textbf{29.53}        &  \textbf{0.8981}      \\
    (b)            &  31.74      & 0.8897      &28.31         &0.7756      & 27.41       &0.7324       &25.51          &0.7669         &29.51          &0.8955       \\ \bottomrule
\end{tabular}}
\end{table*}

\subsection{Effectiveness of  {HASB} Architecture}
To investigate the impact of different configurations of individual modules in  HASB on network performance, we conduct a set of comparative experiments, as shown in Table~\ref{tab:hasb}. For example, on Set5, adding CAB to CB increases the PSNR by 0.09dB and the SSIM by 0.0009. Adding ESA to CB increases the PSNR by 0.14dB and the SSIM by 0.0013. When both modules are added, the PSNR and SSIM increase by 0.2dB and 0.0022, respectively. Compared to the remaining five benchmark datasets, our network achieves the best performance when combining CB with the other two attention modules.

\begin{figure}[!htbp]
\centering
\includegraphics[width=5in]{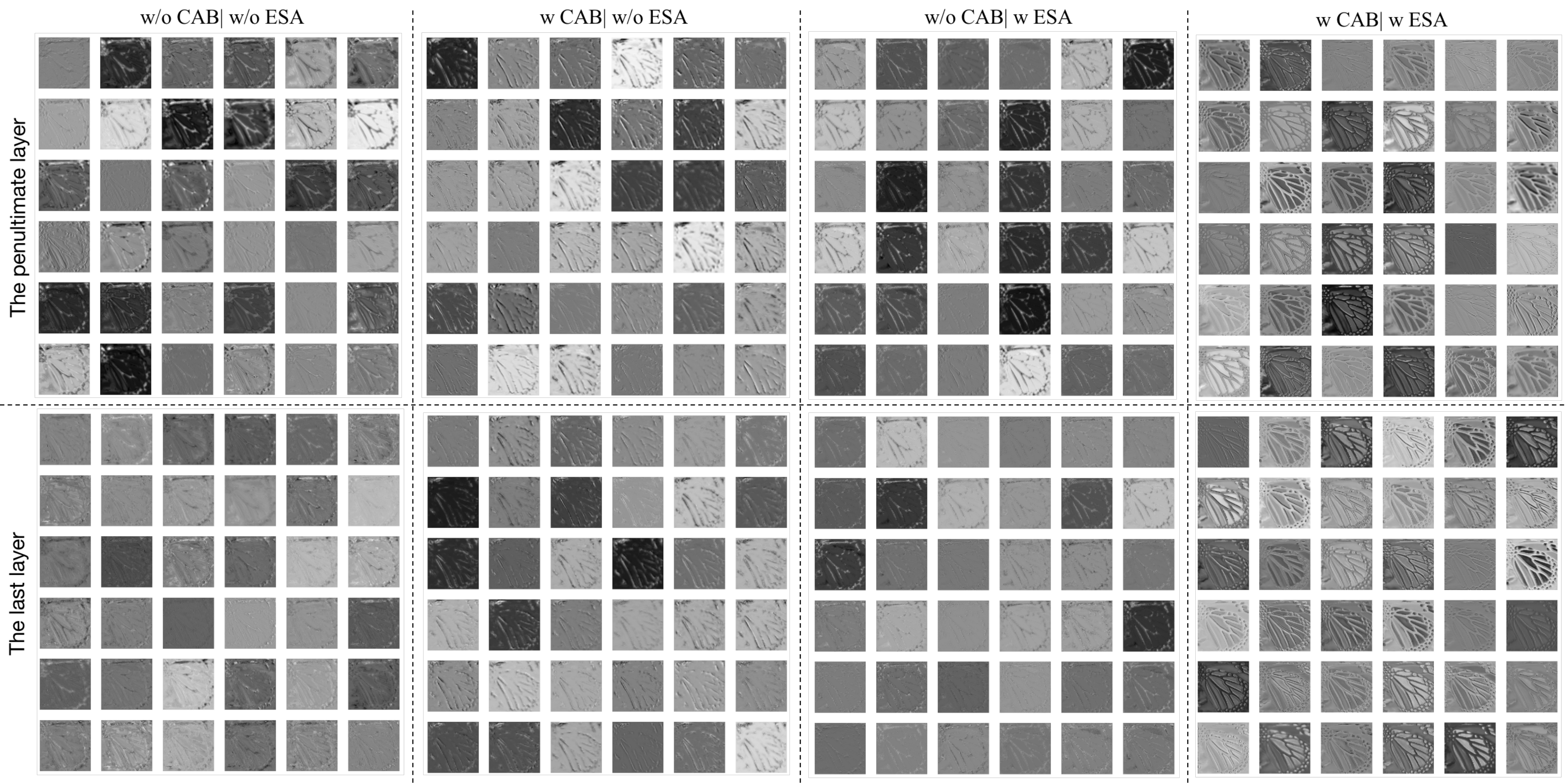}
\caption{Visualization analysis of the impact of CAB and ESA on network feature extraction.}
\label{fig:featuremap}
\end{figure}

To explore the reason behind this phenomenon, we visualize the output features of the last two layers for these four different network structures, as shown in Figure~\ref{fig:featuremap}. We can observe that when these two attention modules are not added, the last two layers of the network extract high-level features that focus on local features with fewer details near the output. In contrast, with the addition of these two attention modules, edges and textures near the network input gradually increase. In low-level vision tasks, low-level features are beneficial for improving network performance. 

Additionally, we aim to investigate the characteristics of HASB in advanced feature extraction and low-level feature retention. Therefore, we select SPAB~\cite{SPAN}, which leverages a parameter-free attention mechanism to achieve feature extraction from shallow to deep layers while maintaining low model complexity and parameter count. We replace HASB with SPAB, keeping all other experimental settings the same.As shown in Table~\ref{tab:span}, the parameter count of HASB is almost half that of SPAB, but it achieves significant improvements in both PSNR and SSIM across five benchmark datasets.

\begin{table*}[!htbp]
\centering
\caption{orange}{Quantitative comparison of SPAB and HASB. The best results are in bold.}
\label{tab:span}
\resizebox{\linewidth}{!}{
\begin{tabular}{cc|cc|cc|cc|cc|cc}
\toprule
\multirow{2}{*}{Method} &    \multirow{2}{*}{Param}    & \multicolumn{2}{c|}{Set5} & \multicolumn{2}{c|}{Set14} & \multicolumn{2}{c|}{B100} & \multicolumn{2}{c|}{Urban100} & \multicolumn{2}{c}{Manga109} \\ 
                              &         & PSNR        & SSIM        & PSNR         & SSIM        & PSNR        & SSIM        & PSNR          & SSIM          & PSNR          & SSIM         \\ \hline
    SPAB             &      836k        &  31.91      & 0.8922      & 28.39        & 0.7776      & 27.45       & 0.7335      & 25.66         & 0.7719        & 29.88       &  0.9004      \\
    HASB            &    435k     & \textbf{32.06}     &   \textbf{0.8937}    &   \textbf{28.52}     &   \textbf{0.7802}   &   \textbf{27.52}    &   \textbf{0.7360}   &     \textbf{25.88}     &   \textbf{0.7798}     &    \textbf{30.12}     &    \textbf{0.9031}        \\ \bottomrule
\end{tabular}}
\end{table*}

\begin{table*}[!htbp]
\centering
\caption{Quantitative results of the state-of-the-art models on five benchmark datasets. The best result is marked with bold. ``CB'' is Convlutional block, which is shown in Figure~\ref{fig:residual_connection}(c).}
\label{tab:hasb}
\resizebox{\linewidth}{!}{
\begin{tabular}{c|c|c|cc|cc|cc|cc|cc}
\toprule
\multirow{2}{*}{Method} & \multirow{2}{*}{ESA} & \multirow{2}{*}{CAB}& \multicolumn{2}{c|}{Set5} & \multicolumn{2}{c|}{Set14} & \multicolumn{2}{c|}{B100} & \multicolumn{2}{c|}{Urban100} & \multicolumn{2}{c}{Manga109} \\ 
                        &   &    & PSNR        & SSIM        & PSNR         & SSIM        & PSNR        & SSIM        & PSNR          & SSIM          & PSNR          & SSIM         \\ \hline

    CB            & \ding{55} & \ding{55} &    31.87    &   0.8915    &    28.38    & 0.7770     &   27.45     &   0.7338   &     25.59     &    0.7700     &     29.53    &   0.8981     \\

     CB        &\ding{55} & \ding{51}&    31.96    &   0.8924    &    28.44    &   0.7787   &    27.48    &   0.7347    &     25.70     &     0.7742    &    29.90     &    0.9005    \\ 

    CB        &\ding{51} & \ding{55}&    32.01    &   0.8928    &    28.42    &   0.7784   &    27.47    &   0.7346   &    25.75      &    0.7757     &     29.86    &    0.9010    \\ 

    CB       &  \ding{51} &  \ding{51} &   \textbf{32.07}     &   \textbf{0.8937}    &   \textbf{28.52}     &   \textbf{0.7802}   &    \textbf{27.52}    &   \textbf{0.7360}   &     \textbf{25.88}     &   \textbf{0.7798}     &    \textbf{30.12}     &    \textbf{0.9031}    \\ 
    
    \bottomrule
\end{tabular}}
\end{table*}

\subsection{Exploration of different activation functions}
Most of the previous SR networks adopt ReLU~\cite{ReLU} or LeakyReLU~\cite{LeakyReLU} as the activation function. ReLU6~\cite{mobilenetv2} is a variant of the ReLU activation function that constrains the output between 0 and 6. It is widely used in mobile and embedded devices because it can provide stable performance in low-precision computing environments.
The results in Table~\ref{tab:act} show that different activation functions can obviously affect the performance of the model. Among these activation functions, ReLU and ReLU6 perform comparably. In our experiments, we chose ReLU6 as the activation function.

\begin{table*}[!htbp]
\centering
\caption{Quantitative comparison of different activation functions. The best result is marked with bold.}
\label{tab:act}
\resizebox{\linewidth}{!}{
\begin{tabular}{c|cc|cc|cc|cc|cc}
\toprule
\multirow{2}{*}{Method} & \multicolumn{2}{c|}{Set5} & \multicolumn{2}{c|}{Set14} & \multicolumn{2}{c|}{B100} & \multicolumn{2}{c|}{Urban100} & \multicolumn{2}{c}{Manga109} \\ 
                               & PSNR        & SSIM        & PSNR         & SSIM        & PSNR        & SSIM        & PSNR          & SSIM          & PSNR          & SSIM         \\ \hline
    ReLU                 &   31.81     &    0.8907   &    \textbf{28.40}    &   \textbf{0.7775}   &   27.45     &   0.7337    &     \textbf{25.60}    &    \textbf{0.7702}    &    \textbf{29.67}    &    \textbf{0.8985}    \\
    LeakyReLU           &    31.80    &    0.8909   &    28.36    &  0.7768    &   27.45     &    0.7336   &     25.60    &    0.7702    &    29.60    &    0.8983    \\ 
    ReLU6           & \textbf{31.87}    &   \textbf{0.8915}    &    28.38    & 0.7770     &   \textbf{27.45}     &   \textbf{0.7338}   &     25.59     &    0.7700     &     29.53    &   0.8981       \\\bottomrule
\end{tabular}}
\end{table*}

\subsection{Effectiveness of Warm-Start Retraining Strategy}
To demonstrate the effectiveness of our proposed warm-start retraining strategy, we use HASN trained from scratch with DIV2K as the baseline. As shown in Table~\ref{tab:wsrs}, when not expanding the training set, our model shows a slight performance improvement with the Warm-Start Retraining Strategy. When further expanding the training set, our model achieves PSNR improvements of 0.11dB, 0.07dB, 0.06dB, 0.15dB, and 0.17dB on the five benchmark datasets. 

\begin{table*}[!htbp]
\centering
\caption{Quantitative comparison of models with and without the Warm-Start Retraining Strategy.``w'' indicates the use of the Warm-Start Retraining Strategy, while ``w/o'' indicates the absence of the Warm-Start Retraining Strategy. The best result is marked with bold.}
\label{tab:wsrs}
\resizebox{\linewidth}{!}{
\begin{tabular}{c|c|cc|cc|cc|cc|cc}
\toprule
\multirow{2}{*}{Method} & \multirow{2}{*}{Dataset} & \multicolumn{2}{c|}{Set5} & \multicolumn{2}{c|}{Set14} & \multicolumn{2}{c|}{B100} & \multicolumn{2}{c|}{Urban100} & \multicolumn{2}{c}{Manga109} \\ 
                           &    & PSNR        & SSIM        & PSNR         & SSIM        & PSNR        & SSIM        & PSNR          & SSIM          & PSNR          & SSIM         \\ \hline

    w/o            &  DIV2k   & {32.06}     &   {0.8937}    &   {28.52}     &   {0.7802}   &   {27.52}    &   {0.7360}   &     {25.88}     &   {0.7798}     &    {30.12}     &    {0.9031}        \\
     w        & DIV2k &    32.08    &    0.8940   &    28.55    &   0.7806    &    27.53    &    0.7365   &     25.89     &     0.7805    &   30.16   &   0.9039     \\

      w        & DF2k &  \textbf{32.17} &   \textbf{0.8953} &  \textbf{28.59}   &\textbf{0.7817}    &  \textbf{27.58}   &   \textbf{0.7377} &  \textbf{26.03}   &   \textbf{0.7846} & \textbf{30.29}    &   \textbf{0.9055} \\
    
    \bottomrule
\end{tabular}}
\end{table*}

\section{Conclusion}
In this paper, we propose a hybrid attention separable network for ffficient image super-resolution(HASN). To make the network more efficient, we use only a few necessary residual connections to avoid gradient vanishing. We design a simple CB module to extract high-level features from the input image and used two essential attention modules (ESA, CAB) to enhance edges and textures near the network input. We conduct extensive feature visualizations to comprehensively analyze the effectiveness of the network structure. Additionally, we propose a warm-start retraining strategy to further exploit the network's performance. Extensive experiments have shown that the proposed method achieves a better balance in performance and lightweight design compared to other networks.

\bibliography{HASN}

\end{document}